# Strain-controlled sign reversal of the anomalous Hall effect in Ru/[Co/Ni]$_N$ multilayers

Jingying Zhang,[1,2,*] Sigang Wang,[1,*] Yue Xiang,[3] Wenhui Xie,[3] Zhe Yuan,[4,†] Yi Liu,[5,6,‡] and Zongzhi Zhang[2,§]

[1]*The School of Physics and Astronomy, Beijing Normal University, Beijing 100875, China.*
[2]*Key Laboratory of Micro and Nano Photonic Structures (MOE), School of Information Science and Technology, Fudan University, Shanghai 200433, China.*
[3]*Engineering Research Center for Nanophotonics and Advanced Instrument, School of Physics and Electronic Science, East China Normal University, Shanghai 200062, China.*
[4]*State Key Laboratory of Surface Physics and Interdisciplinary Center for Theoretical Physics and Information Sciences, Fudan University, Shanghai 200433, China*
[5]*Institute for Quantum Science and Technology, Shanghai University, Shanghai 200444, China*
[6]*Department of Physics, Shanghai University, Shanghai 200444, China*

## Abstract

The anomalous Hall effect (AHE) is a hallmark transport phenomenon in ferromagnets arising from relativistic spin-orbit interaction. Here, we report an unexpected sign reversal of the AHE in Ru/[Co/Ni]$_N$ multilayers controlled by the stacking sequence of the Ru layer. When Ru is placed beneath, rather than atop, the Co/Ni multilayers, the anomalous Hall signal switches from positive to negative. By systematically varying the multilayer repeat number $N$ and combining transport measurements with first-principles calculations, we show that this reversal originates from in-plane tensile strain imposed by the Ru underlayer, which reshapes the electronic structure and redistributes Berry curvature near the Fermi level. Our findings establish interfacial strain as an effective knob for tuning Berry-curvature-driven transport and suggest a pathway toward strain-controlled topological transport phenomena in magnetic multilayers.

**KEYWORDS:** anomalous Hall effect, ferromagnetic multilayers, perpendicular magnetic anisotropy, Berry curvature

[†] Contact author: yuanz@fudan.edu.cn.

[‡] Contact author: yiliu42@shu.edu.cn.

[§] Contact author: zzzhang@fudan.edu.cn.

[*] These authors contributed equally to this work.

# I. INTRODUCTION

The anomalous Hall effect (AHE) is a fundamental spin-orbit-coupling-driven transport phenomenon in ferromagnetic materials (FM), in which a transverse Hall voltage is generated in response to a longitudinal charge current. Beyond serving as a powerful probe of ferromagnetic order and electronic structure [1,2], the AHE has attracted considerable interest for applications in spintronic technologies, including magnetic sensors [3,4], anomalous Hall torque devices [5], and nonvolatile magnetic memory elements [6].

Early experimental and theoretical studies mainly focused on two classes of materials: single-layer films of elemental 3*d* ferromagnetic metals [7-14] and ferromagnetic alloys [12,15,16]. These investigations established the fundamental scaling laws of the AHE and clarified the underlying microscopic mechanisms. In particular, the anomalous Hall response can arise from both extrinsic and intrinsic contributions. The extrinsic mechanisms originate from asymmetric scattering processes, including the skew scattering [17,18] and the side-jump effect [19], whereas the intrinsic contribution arises from the Berry curvature of Bloch electrons in momentum space [20-22].

More recently, increasing attention has been devoted to the anomalous Hall response in ferromagnet/nonmagnet (FM/NM) heterostructures and magnetic multilayers, where interfacial electronic states and spin-orbit coupling can strongly modify transport properties [23-26]. In these systems, the magnitude and even the sign of the AHE can deviate significantly from bulk behavior due to interface-driven effects that remain only partially understood. In particular, polarity reversals of the AHE loop have been observed in perpendicularly magnetized films adjacent to heavy-metal layers [27,28]. Such reversals provide additional degrees of freedom for engineering transport responses and have enabled emerging functionalities in AHE-based spintronic devices [29-31].

Several mechanisms have been proposed to account for these sign reversals. One widely discussed scenario involves competition between bulk and surface (or interface) scattering contributions, which may possess opposite polarities. As the effective dimensionality of a system decreases, surface scattering can become increasingly dominant and eventually reverse the net AHE sign [27,32,33]. Other mechanisms include spin-current transmission and conversion processes at interfaces between ferromagnetic multilayers and adjacent heavy-metal layers [28,34,35], as well as the formation of intermixed alloy layers whose intrinsic AHE polarity differs from that of the host material [36,37]. Despite these proposals, a clear microscopic understanding of the AHE polarity reversal directly linked to electronic structure remains limited. In complex systems such as magnetic multilayers, first-principles calculations are therefore essential for identifying how modifications of band structure and Berry curvature lead to changes in the anomalous Hall response.

In this work, we investigate the AHE in Ru/[Co/Ni]$_N$ heterostructures and reveal an unexpected sign reversal controlled by the stacking sequence. Specifically, the AHE switches from positive to negative when the Ru layer is placed beneath, rather than atop, the Co/Ni multilayers. By systematically varying the repeat number $N$ and combining transport measurements with first-principles calculations, we show that this behavior originates from in-plane tensile strain induced by the Ru underlayer. The strain modifies the electronic structure near the Fermi level and redistributes the Berry curvature, leading to a reversal of the anomalous Hall conductivity (AHC). Our results highlight interfacial strain as an effective route to engineer Berry-curvature-driven transport in metallic multilayer systems.

## II. SAMPLE CHARACTERIZATION

A series of multilayer films with the stack structure Ta (2 nm)/Cu (4 nm)/Ru (0.8 nm)/[Co (0.18 nm)/Ni (0.56 nm)]$_4$ (hereafter denoted as Ru/[Co/Ni]$_4$) were deposited on thermally oxidized Si substrates at room temperature by magnetron sputtering. The base pressure prior to deposition was better than $1\times10^{-8}$ Torr. All thicknesses given in parentheses refer to nominal values.

The sputtering rates for Ta, Cu, Ru, Co, and Ni were calibrated to be 0.76, 1.16, 0.60, 0.47, and 0.50 Å/s, respectively. In this multilayer architecture, the Ta layer serves as a buffer layer to promote film adhesion and smooth growth, while the Cu seed layer is introduced to induce a strong (111) crystallographic texture. Such a texture is known to be essential for stabilizing perpendicular magnetic anisotropy (PMA) in Co/Ni multilayers through interface-driven anisotropy [38]. A thin Ru interlayer was inserted between Cu and the Co/Ni stack to investigate its influence on magnetic and transport properties. Finally, all samples were capped with a sputtered $AlO_x$ layer to prevent surface oxidation.

The crystallographic structure of the films was characterized by X-ray diffraction

(XRD). As shown in Fig. 1(b), a pronounced (111) diffraction peak is observed, confirming the establishment of a strong (111) preferential orientation throughout the multilayer stack.

Magnetic properties were characterized using a vibrating sample magnetometer with magnetic fields applied both in-plane (IP) and out-of-plane (OP) relative to the film plane. Representative hysteresis loops for samples with and without the Ru underlayer are shown in Figs. 1(c) and 1(d), respectively. The square-shaped OP hysteresis loops, together with the large saturation fields observed in the IP configuration, unambiguously indicate robust PMA in both structures. From the hysteresis measurements, the perpendicular coercivity field ( $H_c$ ), effective perpendicular anisotropy field ($H_k$), and saturation magnetization ($M_s$) were extracted. For the [Co/Ni]$_4$ sample, $H_c = 175\ \mathrm{Oe}$, $H_k = 3480\ \mathrm{Oe}$, and $M_s = 686\ \mathrm{emu/cm^3}$, while for the Ru/[Co/Ni]$_4$ sample $H_c = 169\ \mathrm{Oe}$, $H_k = 3930\ \mathrm{Oe}$, and $M_s = 680\ \mathrm{emu/cm^3}$. The close similarity of these parameters indicates that the Ru layer does not significantly modify the static magnetic properties of the [Co/Ni]$_N$ multilayer stack.

For AHE measurements, the films were patterned into Hall bar devices with lateral dimensions of 5 μm × 100 μm using standard ultraviolet photolithography followed by ion-beam etching. Each Hall bar incorporates six electrical contact pads, as schematically illustrated in Fig. 1(a), enabling simultaneous measurements of longitudinal and transverse resistances. Hall resistance measurements were performed by applying a constant current along the longitudinal direction ($x$-axis) while sweeping an external magnetic field along the film normal ($z$-axis) (Supplemental Note 1). The anomalous Hall resistivity, $\rho_{\mathrm{AH}}$, was extracted by linear fitting of the high-field Hall resistance data beyond magnetic saturation.

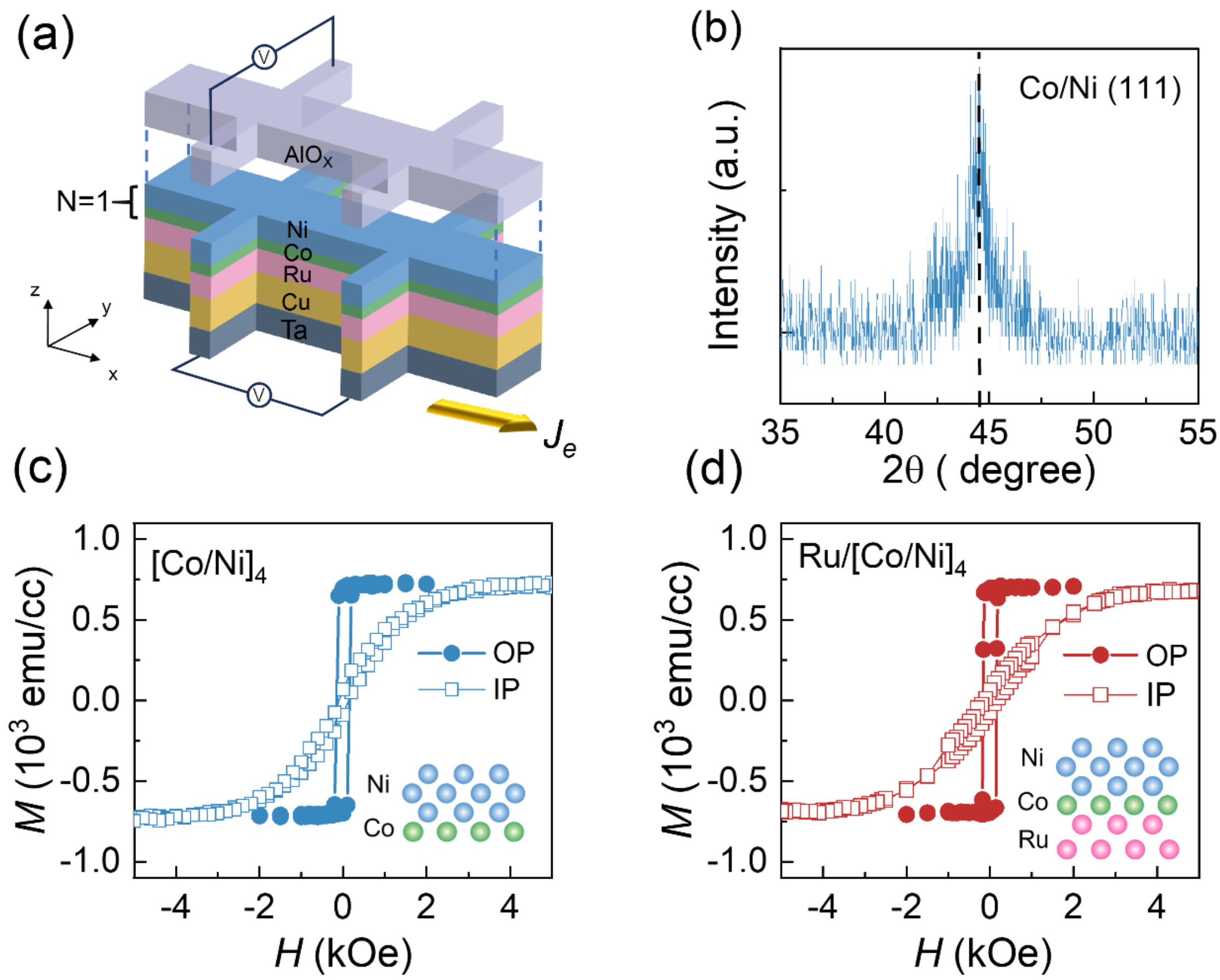

FIG. 1. Structural, magnetic, and transport characterization of Co/Ni multilayers. (a) Schematic illustration of the multilayer stack architecture and Hall bar geometry used for electrical transport measurements. (b) XRD patterns of $[Co/Ni]_N$ multilayers with a repeat number of $N = 16$, showing a pronounced (111) preferential orientation. Out-of-plane and in-plane magnetic hysteresis loops measured for (c) the $[Co/Ni]_4$ and (d) Ru/$[Co/Ni]_4$ samples. The square out-of-plane loops and large in-plane saturation fields confirm robust PMA in both structures. Insets schematically depict the corresponding layer stacking sequences, highlighting the repeated Co/Ni layers and the Ru underlayer.

## III. RESULTS AND DISCUSSION

Figures 2(a)-(d) present the anomalous Hall resistivity ($\rho_{\mathrm{AH}}$) hysteresis loops for $[Co/Ni]_4$ multilayers with different Ru layer configurations. Different from the identical magnetization polarity observed in the out-of-plane hysteresis loops (Figs. 1c, d and Supplemental Note 2), the AHE loop exhibits strikingly different behaviors depending on the presence and position of the Ru layer.

For the reference $[Co/Ni]_4$ multilayer, an anticlockwise AHE loop is observed upon sweeping the out-of-plane magnetic field, as shown in Fig. 2(a), corresponding to a positive anomalous Hall resistivity under positive saturation field. A nearly identical

AHE response is obtained when a Ru layer is deposited on top of the $[Co/Ni]_4$ stack (Fig. 2b), indicating that a Ru capping layer does not measurably influence the anomalous Hall transport. This observation suggests that the dominant contribution to the AHE originates from the Co/Ni multilayer itself and is insensitive to the overlying Ru layer.

In sharp contrast, when the Ru layer is inserted beneath the $[Co/Ni]_4$ multilayer, the AHE loop reverses its polarity and becomes clockwise, accompanied by a pronounced reduction in magnitude, as shown in Fig. 2(c). This behavior indicates a sign reversal of $\rho_{\mathrm{AH}}$, with negative anomalous Hall resistivity under positive saturation field. Notably, this polarity reversal is independent of the interfacial termination of the Co/Ni stack. As shown in Fig. 2(d), reversing the growth sequence to form $Ru/[Ni/Co]_4$, in which Ru is in direct contact with Ni rather than Co, yields essentially the same AHE polarity and comparable magnitude.

The opposite polarities of the AHE loops directly reflect a reversal of the $\rho_{\mathrm{AH}}$ signs. Under sufficiently large positive out-of-plane magnetic fields ($H_z$) that fully saturate the magnetization, $[Co/Ni]_4$ and $[Co/Ni]_4/Ru$ exhibit positive $\rho_{\mathrm{AH}}$ whereas $Ru/[Co/Ni]_4$ and $Ru/[Ni/Co]_4$ display negative $\rho_{\mathrm{AH}}$. This result demonstrates that the overall anomalous Hall response of the Co/Ni multilayer is inverted when the stack is grown on a Ru underlayer. In addition to the sign reversal, the magnitude of $\rho_{AH}$ is substantially suppressed, decreasing from $0.51 \times 10^{-9}\ \Omega \cdot \mathrm{m}$ for $[Co/Ni]_4$ to $-0.16 \times 10^{-9}\ \Omega \cdot \mathrm{m}$ for $Ru/[Co/Ni]_4$.

Importantly, the observed changes in the anomalous Hall response cannot be attributed to trivial current shunting effects. The longitudinal resistance ($R_{xx}$) for the four configurations—$[Co/Ni]_4$, $[Co/Ni]_4/Ru$, $Ru/[Co/Ni]_4$, and $Ru/[Ni/Co]_4$—are 273, 280, 271, and 274 Ω, respectively, indicating comparable current distribution across all samples. This confirms that the Ru-induced sign reversal and suppression of $\rho_{\mathrm{AH}}$ originate from intrinsic modifications to the anomalous Hall transport rather than from geometric or resistive artifacts.

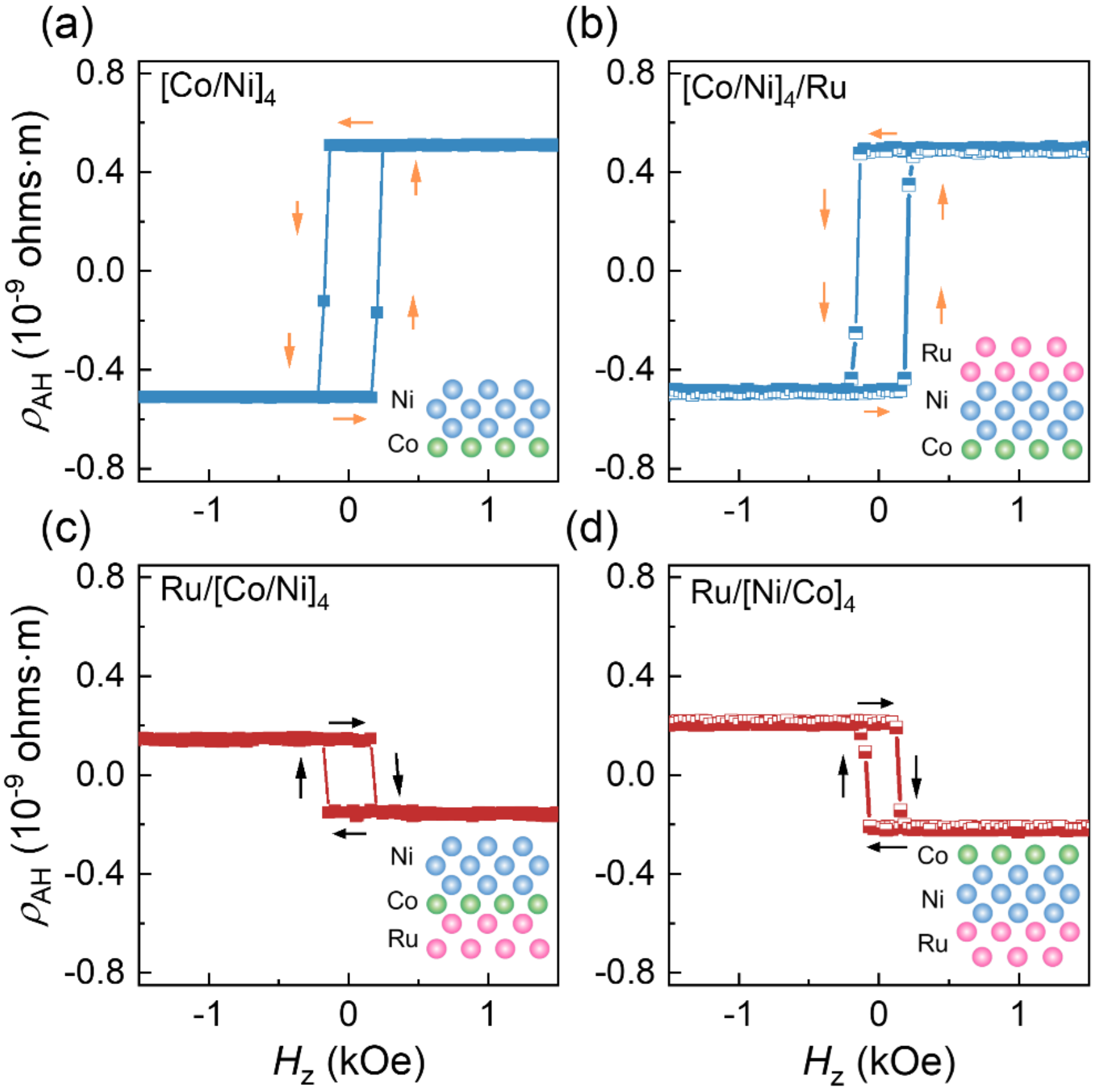

FIG. 2. Ru-layer-induced reversal of the anomalous Hall effect in Co/Ni multilayers. Room temperature anomalous Hall resistivity ($\rho_{\mathrm{AH}}$) as a function of out-of-plane magnetic field ($H_z$) for (a) [Co/Ni]$_4$, (b) [Co/Ni]$_4$/Ru, (c) Ru/[Co/Ni]$_4$, and (d) Ru/[Ni/Co]$_4$ samples. While the [Co/Ni]$_4$ and [Co/Ni]$_4$/Ru structures exhibit identical AHE polarity, the insertion of a Ru underlayer leads to a pronounced reversal and suppression of $\rho_{\mathrm{AH}}$, independent of whether Co or Ni is in direct contact with Ru. Insets schematically illustrate the corresponding layer stacking sequences, highlighting the Co/Ni multilayers and the position of the Ru layer.

Taking the [Co/Ni]$_4$ and Ru/[Co/Ni]$_4$ samples as representative cases with opposite AHE polarities, we next investigate how the number of repeats $N$ in the Co/Ni multilayer influences the anomalous Hall response. The resulting anomalous Hall resistivity $\rho_{\mathrm{AH}}$ as a function of $N$ is summarized in Fig. 3(a). For the [Co/Ni]$_N$ series, $\rho_{\mathrm{AH}}$ decreases monotonically with increasing $N$. This trend correlates well with the simultaneous reduction of the longitudinal resistivity $\rho_{xx}$ (Fig. 3b), which can be attributed to the suppression of finite-size and surface scattering effects as the total multilayer thickness increases [39]. Given the positive correlation between $\rho_{\mathrm{AH}}$ and $\rho_{xx}$ commonly observed in ferromagnetic metals [9,40,41], the monotonic decrease of $\rho_{\mathrm{AH}}$ in the [Co/Ni]$_N$ series is expected.

In contrast, the Ru/[Co/Ni]$_N$ samples exhibit a qualitatively different behavior. While $\rho_{xx}$ still decreases monotonically with increasing $N$, analogous to the Ru-free case, $\rho_{\mathrm{AH}}$ displays a pronounced nonmonotonic dependence on $N$, reaching a minimum negative value at $N$=4. This decoupling between $\rho_{\mathrm{AH}}$ and $\rho_{xx}$ indicates that the anomalous Hall response in the presence of a Ru underlayer cannot be understood solely in terms of conventional resistivity scaling.

To eliminate the influence of longitudinal resistivity and gain insight into the intrinsic anomalous Hall response, we therefore analyze the AHC, defined as $\sigma_{\mathrm{AH}} = \rho_{\mathrm{AH}}/\rho_{xx}^2$. The resulting $\sigma_{\mathrm{AH}}$ values are plotted as a function of $N$ in Fig. 3(c). For the [Co/Ni]$_N$ multilayers, $\sigma_{\mathrm{AH}}$ increases approximately monotonically with $N$, except for the ultrathin case $N = 2$. In this limit, enhanced surface scattering likely introduces additional extrinsic contributions to AHC [14], leading to a slight enhancement relative to the overall trend.

Strikingly, the Ru/[Co/Ni]$_N$ series retains a strongly nonmonotonic behavior in $\sigma_{\mathrm{AH}}$, including sign reversals as a function of $N$. Considering that bulk Co and Ni possess anomalous Hall conductivities of opposite sign—approximately 480 S/cm for Co [42] and $-646$ S/cm for Ni [43]—the observed sign change in Ru/[Co/Ni]$_N$ suggests that the multilayer can exhibit either Co-like or Ni-like anomalous Hall behavior depending on the repeat number. By contrast, the Ru-free [Co/Ni]$_N$ multilayers remain dominantly Co-like over the entire thickness range investigated.

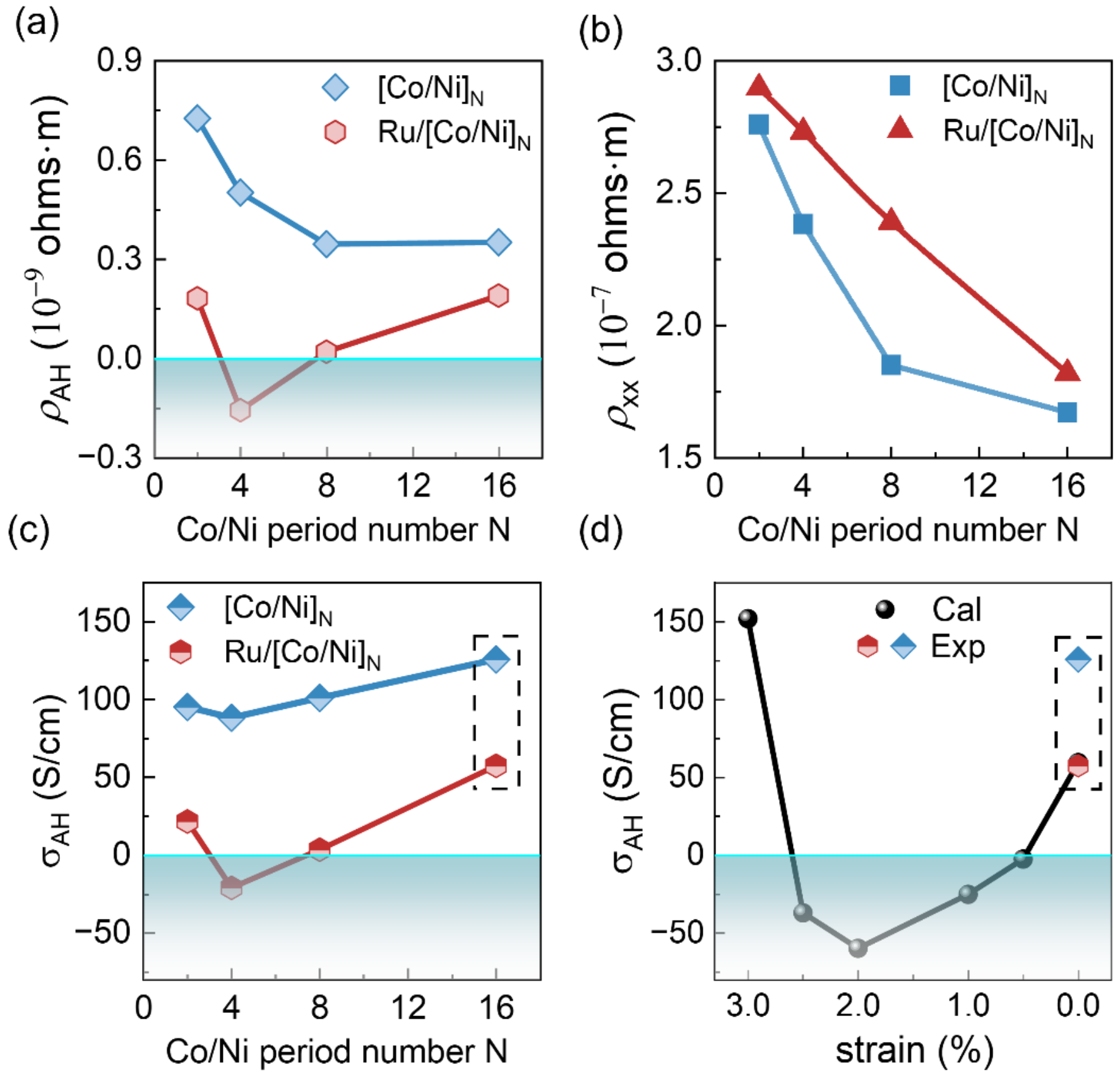

FIG. 3. Thickness-dependent anomalous Hall transport and strain-induced sign reversal. (a) Anomalous Hall resistivity $\rho_{\mathrm{AH}}$, (b) longitudinal resistivity $\rho_{xx}$ and (c) AHC $\sigma_{\mathrm{AH}}$ as a function of the Co/Ni repeat number $N$ for $[\mathrm{Co/Ni}]_N$ and $\mathrm{Ru/[Co/Ni]}_N$ multilayers. While the Ru-free $[\mathrm{Co/Ni}]_N$ samples exhibit a largely monotonic evolution of $\sigma_{\mathrm{AH}}$, the $\mathrm{Ru/[Co/Ni]}_N$ series displays pronounced nonmonotonic behavior and sign reversals. (d) Calculated $\sigma_{\mathrm{AH}}$ for Co/Ni multilayer as a function of in-plane strain, demonstrating a strain-driven sign change. The colored data points highlighted by dashed boxes in (c) and (d) correspond to the same experimental values, illustrating the consistency between experiment and theory.

To elucidate the physical mechanism underlying the observed Ru-induced reversal of the AHE, we performed first-principles calculations of the AHC for bulk Co/Ni multilayers. Computational details are provided in Supplementary Notes S3. The calculated bulk AHC value of 59.5 S/cm is in reasonable agreement with the experimentally measured values obtained from thick samples, namely 125.8 S/cm for $[\mathrm{Co/Ni}]_{16}$ and 57.5 S/cm for $\mathrm{Ru/[Co/Ni]}_{16}$ (Fig. 3d). This agreement indicates that, in sufficiently thick Co/Ni multilayers, the anomalous Hall response is predominantly governed by the intrinsic bulk electronic structure, with minimal influence from the underlying Ru layer.

The in-plane lattice constant of hexagonal Ru is 2.70 Å, which is approximately 8% larger than that of the Co/Ni(111) multilayer. As a consequence of this lattice mismatch, thin Co/Ni multilayers grown on Ru are expected to experience significant tensile in-plane strain at the interface. In magnetic thin films, such strain is known to strongly modify electronic and magnetic properties [44-46], until it gradually relaxes with increasing film thickness. To assess the impact of lattice strain on the anomalous Hall response, we calculated the AHC $\sigma_{\mathrm{AH}}$ of Co/Ni multilayers under various levels of in-plane strain in the (111) plane, while conserving the atomic volume. The calculated results are shown by the black symbols in Fig. 3(d). As the in-plane lattice constant increases, $\sigma_{\mathrm{AH}}$ exhibits a pronounced nonmonotonic dependence and undergoes two sign reversals. This behavior qualitatively reproduces the experimentally observed thickness dependence of $\sigma_{\mathrm{AH}}$ in Ru/[Co/Ni]$_N$ multilayers, strongly suggesting that the anomalous Hall sign reversal originates from Ru-induced lattice strain.

In the absence of a Ru underlayer, as in the [Co/Ni]$_N$ samples (Fig. 2a), or when Ru is deposited on top of the Co/Ni multilayers (Fig. 2b), the lattice strain exerted by the Cu seed layer is relatively weak, and the AHE polarity remains unchanged. By contrast, when Ru is placed beneath the Co/Ni stack, both Co and Ni layers—whose in-plane lattice constants are smaller than that of Ru—are stretched regardless of which element is in direct contact with Ru. The resulting tensile strain leads to the experimentally observed reversal of the anomalous Hall polarity in Ru/[Co/Ni]$_N$ and Ru/[Ni/Co]$_N$ (Fig. 2c,d), consistent with the strain-dependent theoretical predictions.

To identify the microscopic origin of the strain-induced sign change in $\sigma_{\mathrm{AH}}$, we performed a comparative analysis between the unstrained (0%) and tensile-strained (+2%) Co/Ni multilayers. The corresponding bulk Brillouin zones (BZ) are nearly identical in shape for the two cases, as shown in Fig. 4(a). Since the AHC is obtained by integrating the Berry curvature $\Omega_{xy}$ over the entire BZ, we decompose $\sigma_{\mathrm{AH}}$ into contributions from individual $k_z$ planes. The resulting $k_z$-resolved $\sigma_{\mathrm{AH}}(k_z)$ is plotted in Fig. 4(b). Both unstrained and strained structures exhibit negative contributions near the BZ center and positive contributions for $k_z > 0.6\pi/c$, where $c$ is the lattice constant perpendicular to the multilayer plane. The most pronounced difference between the two cases occurs around $k_z = 0.5\pi/c$, where the contributions exhibit opposite signs. Integrating over $k_z$ reproduces the bulk $\sigma_{\mathrm{AH}}$ values (inset of Fig. 4b), confirming that the sign reversal is dominated by electronic states near this $k_z$ plane.

We therefore calculated the $k$-resolved Berry curvature at $k_z = 0.5\pi/c$ and plotted it in a relative form defined $-\Omega^{\mathrm{r}}_{xy} = -\mathrm{sgn}(\Omega_{xy}) \log_{10}|\Omega_{xy}|$, as shown in Figs. 4(c) and 4(d). The minus sign ensures that the Berry curvature retains the same sign convention as $\sigma_{\mathrm{AH}}$. For the unstrained lattice, $-\Omega^{\mathrm{r}}_{xy}$ is predominantly positive, whereas for the strained lattice it is largely negative, with pronounced hot spots along the $L' - K'$ direction in both cases.

To identify the electronic origin of these hot spots, we further analyzed the band structure and Berry curvature along the $L' - K'$ direction (Figs. 4e,f), projecting each band onto Co- and Ni-derived components (red and blue, respectively). Near the Fermi level, the electronic structure is dominated by Ni bands (spin-down), accompanied by a single Co band of the same spin. In the unstrained lattice (Fig. 4e), a gap at the Fermi level forms between the Co-dominated band and strongly hybridized Co-Ni bands, giving rise to a Berry curvature peak. Because Co intrinsically exhibits a positive AHC, this Co-dominant gap yields a positive contribution to $\sigma_{\mathrm{AH}}$.

Under tensile strain (Fig. 4f), the Ni-dominated bands are shifted upward toward the Fermi level, resulting in the formation of two gaps that generate pronounced Berry curvature peaks. One gap emerges near the $L'$ point due to strain and is formed predominantly by Ni-derived bands, while the other gap—corresponding to that in the unstrained case—is now formed between a Co band and a Ni-dominated band. Given the intrinsically negative AHC of Ni, both gaps contribute negatively to $\sigma_{\mathrm{AH}}$. Further orbital analysis (not shown) reveals that the relevant Ni-dominated band primarily originates from the $d_{z^2}$ orbital of spin-down Ni electrons.

The strain-induced sign reversal of the AHC can therefore be understood as follows. In Co/Ni multilayers, the number of Ni spin-down bands below the Fermi level exceeds that of Co. Tensile in-plane strain reduces the interlayer spacing along the (111) direction, enhances interlayer hybridization, and shifts Ni-derived bands upward in energy, leading to strain-induced gap openings at the Fermi level. These gaps produce dominant negative Berry curvature contributions. In the experimental samples, such tensile strain arises from the large lattice mismatch between Ru and Co/Ni layers, persists in thin multilayers, and gradually relaxes as the magnetic thickness increases [44,46]. Consequently, a transition from positive to negative AHC is observed as the Co/Ni multilayer thickness decreases, as shown in Fig. 3(c).

It is worth mentioning that there could be a distinction between the underlying mechanisms in the ultrathin sample ($N = 2$) and the high-strain bulk calculations, which account for the second AHC sign reversal, *i.e.*, the transition from negative to positive. For the experimental $N = 2$ sample, the total magnetic film thickness is only ≈ 1.5 nm. In this ultrathin limit, extrinsic boundary scattering from the capping ($AlO_x$) and buffer (Ru/Cu) interfaces, alongside dimensional confinement, becomes prominent. In contrast, in the calculated bulk system [Fig. 3(d)], a second sign reversal occurs at strains exceeding 2.0% presumably due to intrinsic strain-induced shifts of Ni $d$-bands farther across $E_F$. Consequently, the deviation at $N = 2$ in experiment reflects enhanced surface/interface scattering contributions, whereas the high-strain trend in DFT captures the intrinsic bulk topological evolution.

We also explicitly address the potential influence of interfacial atomic intermixing. Given the sub-nanometer thickness of the Co layer (nominally 0.18 nm) and the large lattice mismatch with the Ru underlayer, a degree of Co-Ni intermixing is expected during deposition. However, quantitative considerations indicate that intermixing cannot account for the observed AHE sign reversal. Both theoretical calculations [12] and experimental measurements [47] on unstrained binary $Co_{1-x}Ni_x$ alloys show that a negative AHC requires a high Ni atomic concentration exceeding 80% ($x \geq 0.8$). For our nominal [Co (0.18 nm)/Ni (0.56 nm)]$_N$ multilayers, the average Ni fraction is ≈ 75%, and local intermixing near the Co layers yields an even lower local Ni concentration. Consequently, unstrained interfacial intermixing cannot cross the concentration threshold required to reverse the AHC sign. Furthermore, our DFT calculations—based on ideal strained superlattices without intermixing—reproduce the sign reversal under tensile strain [Fig. 3(d)]. This confirms that lattice-strain-induced Berry curvature redistribution, rather than interfacial intermixing, is the dominant mechanism governing the topological transport reversal.

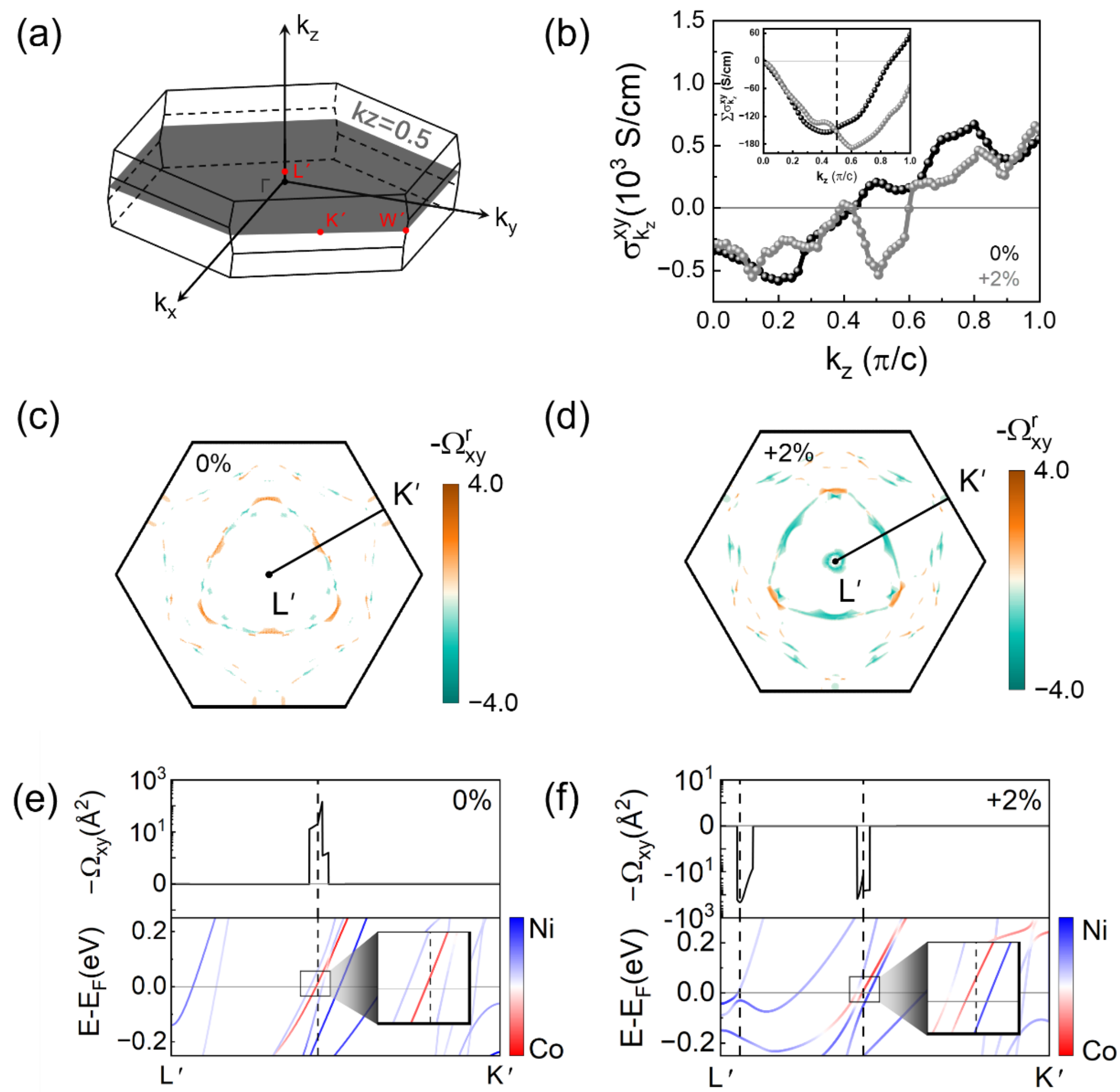

FIG. 4. Strain-induced redistribution of Berry curvature and AHC of the Co/Ni multilayer. (a) BZ of the Co/Ni multilayer, indicating the high-symmetry points and the $L' - K'$ path analyzed. (b) Calculated $k_z$-resolved AHC $\sigma_{\mathrm{AH}}(k_z)$ for Co/Ni multilayers with unstrained and tensile-strained lattices. The inset shows the cumulative integral of $\sigma_{\mathrm{AH}}(k_z)$ from $-k_z$ to $k_z$, illustrating how the bulk AHC emerges from contributions of specific $k_z$ planes. Distribution of the Berry curvature in the $k_z = 0.5\pi/c$ plane for the (c) unstrained and (d) strained Co/Ni multilayers, plotted in a relative representation (see text). The sign reversal of the dominant Berry-curvature contributions under strain is evident. Berry curvature (upper panels) and corresponding electronic band structure near the Fermi level (lower panels) along the $L' - K'$ direction for the (e) unstrained and (f) strained lattices. The strain-induced evolution of Co-Ni hybridized bands gives rise to distinct Berry-curvature hot spots that drive the sign change of the AHC.

## IV. CONCLUSIONS

In conclusion, we demonstrate that the anomalous Hall effect in Co/Ni multilayers can be reversibly tuned in both magnitude and sign through interfacial strain engineering, without altering the overall magnetic ground state. By introducing a Ru underlayer with a large lattice mismatch, tensile in-plane strain is imposed on the Co/Ni multilayers, leading to a redistribution of Berry curvature and a strain-driven inversion of the AHC. First-principles calculations reveal that this effect originates from strain-induced modifications of Co-Ni hybridized electronic states near the Fermi level, which selectively enhance Ni-dominated contributions with opposite topological character. Our results establish lattice strain as a powerful and versatile control parameter for topological transport in metallic ferromagnets, extending beyond conventional thickness or composition tuning. More broadly, this work highlights the potential of interface-induced strain to engineer Berry-curvature-driven phenomena in multilayer heterostructures, opening new pathways for designing functional spintronic devices based on controllable topological responses.

**Acknowledgments**

This work is supported by the National Key Research and Development Program of China (Grant No. 2024YFA1408500) and the National Natural Science Foundation of China (Grants No. 12374104, No. 52171230, No. 12374101, and No. 12574115).

## References

[1] N. Nagaosa, J. Sinova, S. Onoda, A. H. MacDonald, and N. P. Ong, Anomalous Hall effect, Rev. Mod. Phys. **82**, 1539 (2010).
[2] D. Culcer, in *Encyclopedia of Condensed Matter Physics (Second Edition)*, edited by T. Chakraborty (Academic Press, Oxford, 2024), pp. 587.
[3] K. Wang, Y. Zhang, and G. Xiao, Anomalous Hall Sensors with High Sensitivity and Stability Based on Interlayer Exchange-Coupled Magnetic Thin Films, Phys. Rev. Appl. **13** (2020).
[4] W. Su, Z. Hu, Y. Li, Y. Han, Y. Chen, C. Wang, Z. Jiang, Z. He, J. Wu, Z. Zhou *et al.*, Easy-Cone Magnetic State Induced Ultrahigh Sensitivity and Low Driving Current in Spin-Orbit Coupling 3D Magnetic Sensors, Adv. Funct. Mater. **33** (2022).
[5] E. A. Montoya, X. Pei, and I. N. Krivorotov, Anomalous Hall spin current drives self-generated spin-orbit torque in a ferromagnet, Nat. Nanotechnol. **20**, 353 (2025).
[6] Z. Feng, H. Yan, X. Wang, H. Guo, P. Qin, X. Zhou, Z. Chen, H. Wang, Z. Jiao, Z. Leng *et al.*, Nonvolatile Electric Control of the Anomalous Hall Effect in an Ultrathin Magnetic Metal, Adv. Electron. Mater. **6**, 1901084 (2019).
[7] Y. Yao, L. Kleinman, A. H. MacDonald, J. Sinova, T. Jungwirth, D.-s. Wang, E. Wang, and Q. Niu, First Principles Calculation of Anomalous Hall Conductivity in Ferromagnetic bcc Fe, Phys. Rev. Lett. **92**, 037204 (2004).
[8] X. Wang, D. Vanderbilt, J. R. Yates, and I. Souza, Fermi-surface calculation of the anomalous Hall conductivity, Phys. Rev. B **76**, 195109 (2007).
[9] Y. Tian, L. Ye, and X. Jin, Proper scaling of the anomalous Hall effect, Phys. Rev. Lett. **103**, 087206 (2009).
[10] H.-R. Fuh and G.-Y. Guo, Intrinsic anomalous Hall effect in nickel: A GGA+U

study, Phys. Rev. B **84**, 144427 (2011).
[11] L. Ye, Y. Tian, X. Jin, and D. Xiao, Temperature dependence of the intrinsic anomalous Hall effect in nickel, Phys. Rev. B **85**, 220403 (2012).
[12] I. Turek, J. Kudrnovský, and V. Drchal, *Ab initio* theory of galvanomagnetic phenomena in ferromagnetic metals and disordered alloys, Phys. Rev. B **86**, 014405 (2012).
[13] J.-C. Tung, H.-R. Fuh, and G.-Y. Guo, Anomalous and spin Hall effects in hcp cobalt from GGA+U calculations, Phys. Rev. B **86**, 024435 (2012).
[14] V. L. Grigoryan, J. Xiao, X. Wang, and K. Xia, Anomalous Hall effect scaling in ferromagnetic thin films, Phys. Rev. B **96**, 144426 (2017).
[15] S. Mankovsky, S. Wimmer, S. Polesya, and H. Ebert, Composition-dependent magnetic response properties of $Mn_{1-x}Fe_xGe$ alloys, Phys. Rev. B **97**, 024403 (2018).
[16] O. Šipr, S. Wimmer, S. Mankovsky, and H. Ebert, Transport properties of doped permalloy via *ab initio* calculations: Effect of host disorder, Phys. Rev. B **101**, 085109 (2020).
[17] J. Smit, The spontaneous Hall effect in ferromagnetics I, Physica **21**, 877 (1955).
[18] J. Smit, The spontaneous hall effect in ferromagnetics II, Physica **24**, 39 (1958).
[19] L. Berger, Side-jump mechanism for the Hall effect of ferromagnets, Phys. Rev. B **2**, 4559 (1970).
[20] R. Karplus and J. M. Luttinger, Hall Effect in Ferromagnetics, Phys. Rev. **95**, 1154 (1954).
[21] T. Jungwirth, Q. Niu, and A. H. MacDonald, Anomalous Hall effect in ferromagnetic semiconductors, Phys. Rev. Lett. **88**, 207208 (2002).
[22] M. Onoda and N. Nagaosa, Topological Nature of Anomalous Hall Effect in Ferromagnets, J. Phys. Soc. Jpn. **71**, 19 (2002).
[23] S.-L. Jiang, X. Chen, X.-J. Li, K. Yang, J.-Y. Zhang, G. Yang, Y.-W. Liu, J.-H. Lu, D.-W. Wang, J. Teng *et al.*, Anomalous Hall effect engineering via interface modification in Co/Pt multilayers, Appl. Phys. Lett. **107**, 112404 (2015).
[24] Hellman F, A. Hoffmann, Y. Tserkovnyak, G. S. D. Beach, E. E. Fullerton, C. Leighton, A. H. MacDonald, D. C. Ralph, D. A. Arena, H. A. Durr *et al.*, Interface-induced phenomena in magnetism, Rev. Mod. Phys. **89**, 026006 (2017).
[25] W. L. Peng, J. Y. Zhang, Y. W. Liu, G. N. Feng, L. Wang, and G. H. Yu, Tunable anomalous Hall effect in multilayers induced by artificial interfacial scattering dots, AIP Adv. **8**, 035206 (2018).
[26] J. Zhang, W. Peng, G. Yu, Z. He, F. Yang, W. Ji, C. Hu, and S. Wang, Tunable Giant Anomalous Hall Angle in Perpendicular Multilayers by Interfacial Orbital Hybridization, ACS Appl. Mater. Interfaces **11**, 24751 (2019).
[27] D. Rosenblatt, M. Karpovski, and A. Gerber, Reversal of the extraordinary Hall effect polarity in thin Co/Pd multilayers, Appl. Phys. Lett. **96**, 022512 (2010).
[28] T. H. Dang, Q. Barbedienne, D. Q. To, E. Rongione, N. Reyren, F. Godel, S. Collin, J. M. George, and H. Jaffrès, Anomalous Hall effect in 3*d*/5*d* multilayers mediated by interface scattering and nonlocal spin conductivity, Phys. Rev. B **102**, 144405 (2020).
[29] T. Taniguchi, J. Grollier, and M. D. Stiles, Spin-Transfer Torques Generated by the Anomalous Hall Effect and Anisotropic Magnetoresistance, Phys. Rev. Appl. **3** (2015).
[30] B. Dieny, I. L. Prejbeanu, K. Garello, P. Gambardella, P. Freitas, R. Lehndorff, W. Raberg, U. Ebels, S. O. Demokritov, J. Akerman *et al.*, Opportunities and challenges for spintronics in the microelectronics industry, Nat. Electron. **3**, 446 (2020).
[31] D.-F. Shao, S.-H. Zhang, R.-C. Xiao, Z.-A. Wang, W. J. Lu, Y. P. Sun, and E. Y. Tsymbal, Spin-neutral tunneling anomalous Hall effect, Phys. Rev. B **106**, L180404

(2022).
[32] Z. B. Guo, W. B. Mi, R. O. Aboljadayel, B. Zhang, Q. Zhang, P. G. Barba, A. Manchon, and X. X. Zhang, Effects of surface and interface scattering on anomalous Hall effect in Co/Pd multilayers, Phys. Rev. B **86**, 104433 (2012).
[33] N. Jiang, B. Yang, Y. Bai, Y. Jiang, and S. Zhao, The sign reversal of anomalous Hall effect derived from the transformation of scattering effect in cluster-assembled Ni(0.8)Fe(0.2) nanostructural films, Nanoscale **13**, 11817 (2021).
[34] N. Song, Y. X. Huang, K. Ren, J. X. Ding, T. Li, J. K. Zhang, J. L. Xie, Y. H. Xu, Z. H. Ding, J. S. Zhu *et al.*, The influence of orbital moments in anomalous Hall effect in Co/Ni multilayers with perpendicular magnetic anisotropy, AIP Adv. **12**, 025316 (2022).
[35] X. Shi, Q. Ping, X. Zhang, K. Xiao, P. Duan, L. Du, and W. Mi, Unveiling the role of temperature in sign reversal of anomalous Hall resistivity in epitaxial Pt(3 nm)/$Fe_4N$(≤6 nm)/MgO heterostructures toward spintronic devices, Appl. Phys. Lett. **126** (2025).
[36] Y. Zhang and S. Granville, Two-channel anomalous Hall effect originating from the intermixing in $Mn_2CoAl$/Pd thin films, Phys. Rev. B **106**, 144414 (2022).
[37] Y. Zhang, D. Cortie, T. LaGrange, W. Lee, T. Butler, B. Ludbrook, and S. Granville, Unraveling the sign reversal of the anomalous Hall effect in ferromagnet/heavy-metal ultrathin films, Phys. Rev. B **107**, 094408 (2023).
[38] G. H. O. Daalderop, P. J. Kelly, and F. J. A. den Broeder, Prediction and confirmation of perpendicular magnetic anisotropy in Co/Ni multilayers, Phys. Rev. Lett. **68**, 682 (1992).
[39] Y. Yang, Z. Luo, H. Wu, Y. Xu, R. W. Li, S. J. Pennycook, S. Zhang, and Y. Wu, Anomalous Hall magnetoresistance in a ferromagnet, Nat. Commun. **9**, 2255 (2018).
[40] H. Hayashi, A. Asami, and K. Ando, Anomalous Hall effect at a $PtO_x$/Co interface, Phys. Rev. B **100**, 214415 (2019).
[41] M.-W. Yoo, J. Tornos, A. Sander, L.-F. Lin, N. Mohanta, A. Peralta, D. Sanchez-Manzano, F. Gallego, D. Haskel, J. W. Freeland *et al.*, Large intrinsic anomalous Hall effect in $SrIrO_3$ induced by magnetic proximity effect, Nat. Commun. **12** (2021).
[42] T. Miyasato, N. Abe, T. Fujii, A. Asamitsu, S. Onoda, Y. Onose, N. Nagaosa, and Y. Tokura, Crossover behavior of the anomalous Hall effect and anomalous nernst effect in itinerant ferromagnets, Phys. Rev. Lett. **99**, 086602 (2007).
[43] J. M. Lavine, Extraordinary Hall-Effect Measurements on Ni, Some Ni Alloys, and Ferrites, Phys. Rev. **123**, 1273 (1961).
[44] S. Tan, Y. Zhang, M. Xia, Z. Ye, F. Chen, X. Xie, R. Peng, D. Xu, Q. Fan, H. Xu *et al.*, Interface-induced superconductivity and strain-dependent spin density waves in FeSe/$SrTiO_3$ thin films, Nat. Mater. **12**, 634 (2013).
[45] D. Tian, Z. Liu, S. Shen, Z. Li, Y. Zhou, H. Liu, H. Chen, and P. Yu, Manipulating Berry curvature of $SrRuO_3$ thin films via epitaxial strain, Proc. Natl. Acad. Sci. U.S.A. **118**, e2101946118 (2021).
[46] H. Chi, Y. Ou, T. B. Eldred, W. Gao, S. Kwon, J. Murray, M. Dreyer, R. E. Butera, A. C. Foucher, H. Ambaye *et al.*, Strain-tunable Berry curvature in quasi-two-dimensional chromium telluride, Nat. Commun. **14**, 3222 (2023).
[47] W. J. Fan, L. Ma, and S. M. Zhou, Sign change of skew scattering induced anomalous Hall conductivity in epitaxial NiCo(002) films: band filling effect, Journal of Physics D: Applied Physics **48** (2015).

# Supplemental Material for "Strain-controlled sign reversal of the anomalous Hall effect in Ru/[Co/Ni]$_N$ multilayers"

Jingying Zhang,[1,2,*] Sigang Wang,[1,*] Yue Xiang,[3] Wenhui Xie,[3] Zhe Yuan,[4,†] Yi Liu,[5,6,‡] and Zongzhi Zhang[2,§]

*[1]The School of Physics and Astronomy, Beijing Normal University, Beijing 100875, China.*

*[2]Key Laboratory of Micro and Nano Photonic Structures (MOE), School of Information Science and Technology, Fudan University, Shanghai 200433, China.*

*[3]Engineering Research Center for Nanophotonics and Advanced Instrument, School of Physics and Electronic Science, East China Normal University, Shanghai 200062, China.*

*[4]State Key Laboratory of Surface Physics and Interdisciplinary Center for Theoretical Physics and Information Sciences, Fudan University, Shanghai 200433, China*

*[5]Institute for Quantum Science and Technology, Shanghai University, Shanghai 200444, China*

*[6]Department of Physics, Shanghai University, Shanghai 200444, China*

## 1. Measurement of $R_{AH}$

We measured the Hall resistance of the $[Co/Ni]_4$ and Ru/$[Co/Ni]_4$ multilayers at room temperature. A constant current was applied along the longitudinal direction (*x*-axis) while sweeping an external magnetic field along the film normal (*z*-axis), as illustrated in Fig. 1(a). Fig. S1 shows the resulting anomalous Hall effect loops, which exhibit opposite polarities for the two sample configurations. To determine the anomalous Hall resistivity, $\rho_{\mathrm{AH}}$, we extracted the high-field Hall resistance data beyond magnetic saturation through linear fitting.

[†] Contact author: yuanz@fudan.edu.cn.

[‡] Contact author: yiliu42@shu.edu.cn.

[§] Contact author: zzzhang@fudan.edu.cn.

* These authors contributed equally to this work.

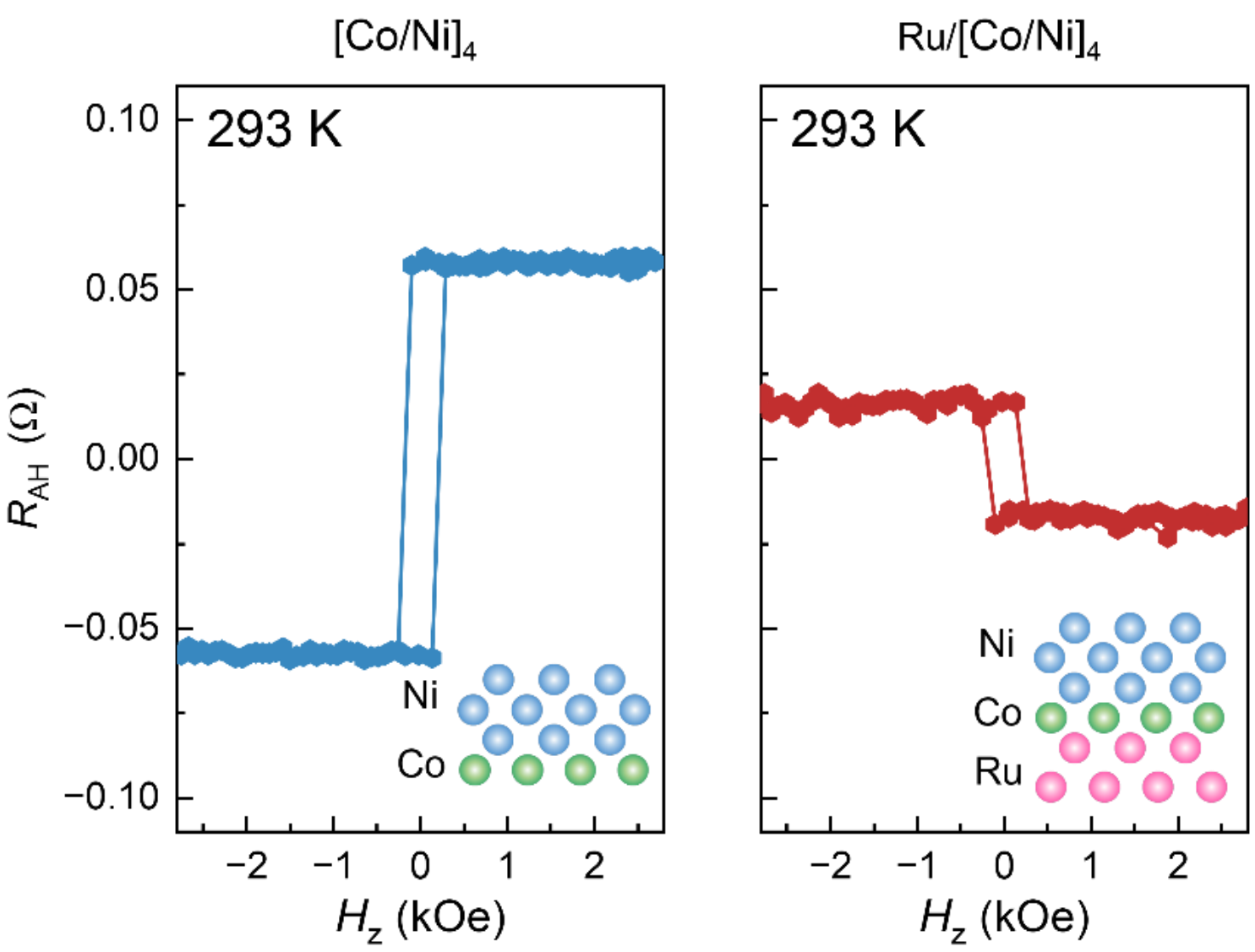

Fig. S1. Anomalous Hall effect loops for (a) the $[Co/Ni]_4$ and (b) $Ru/[Co/Ni]_4$ samples measured at 293 K. The insets schematically depict the corresponding layer stacking sequences, highlighting the repeated Co/Ni layers and the Ru underlayer.

## 2. The magnetic hysteresis loops

Magnetic hysteresis loops for the $[Co/Ni]_4/Ru$ stack (Ru deposited on top) were obtained using both in-plane (IP) and out-of-plane (OP) magnetic fields, as shown in Fig. S2(a). Figure S2(b) displays the corresponding loops for the $Ru/[Ni/Co]_4$ sample, where the Ru layer serves as the underlayer and the growth sequence is reversed (Ni is in direct contact with Ru). Both structures exhibit square OP loops and large IP saturation fields, confirming robust perpendicular magnetic anisotropy. The saturation magnetization magnitudes of these two samples are nearly identical to those of the $[Co/Ni]_4$ and $Ru/[Co/Ni]_4$ samples (shown in Figs. 1c,d), indicating that the position of the Ru layer does not significantly influence the magnetic properties.

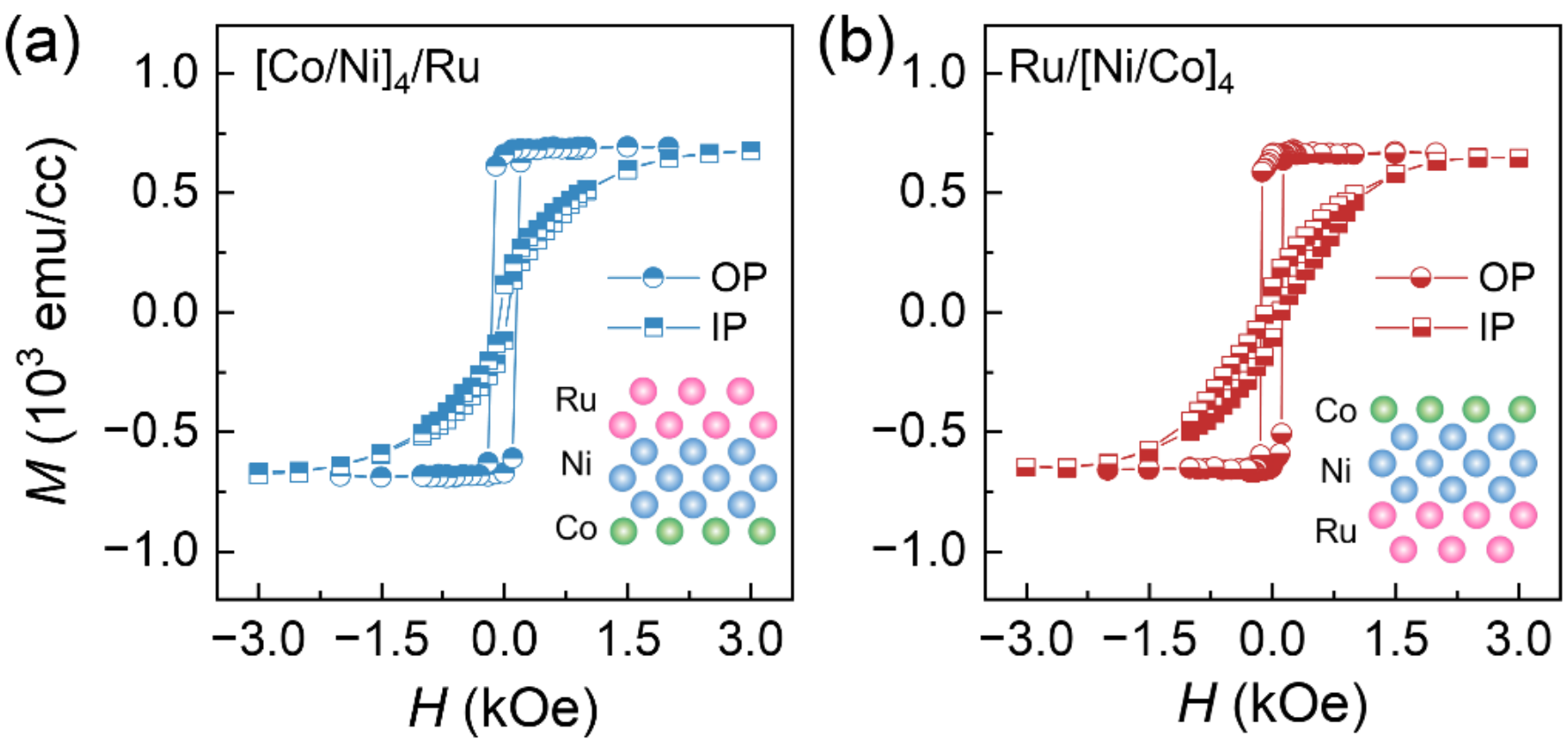

Fig. S2. Out-of-plane and in-plane magnetic hysteresis loops for the (a) $[Co/Ni]_4/Ru$ and (b) $Ru/[Ni/Co]_4$ samples. Insets schematically illustrate the corresponding layer stacking sequences, highlighting the Co/Ni multilayers and the specific position of the Ru layer.

## 3. Computational details

Based on the experimental stack structure, we performed first-principles calculations of the anomalous Hall conductivity for $Co_1Ni_3$. The electronic structure was investigated using the Vienna *ab initio* simulation package (VASP) [1]. We employed the Perdew-Burke-Ernzerhof (PBE) exchange-correlation functional within the GGA+U approach [2]. To ensure accurate electronic structure results, a U value of 4.8 eV was used for Ni [3]. The plane-wave cutoff energy was set to 400 eV, with energy and force convergence criteria of $10^{-6}$ eV and 0.001 eV/Å, respectively. Brillouin zone sampling was performed using a Γ-centered 22×22×22 $k$-point mesh. The anomalous Hall conductivity and Berry curvature were calculated using the Wannier90 [4] software packages.

In this work, the lattice constant of the unit cell for face-centered-cubic (fcc) $Co_1Ni_3$ is $a_0 = 3.526$ Å, determined via the Vegard's law using the lattice constants of fcc Ni (3.524 Å) and hexagonal-close-packed Co ($a = 2.507$ Å and $c/a = 1.623$). Strain was applied while maintaining a constant unit cell volume under in-plane isotropic strain, defined as $\varepsilon = \left(a_{\mathrm{strain}}^{\mathrm{ip}} - a_0^{\mathrm{ip}}\right)/a_0^{\mathrm{ip}} = \Delta a^{\mathrm{ip}}/a_0^{\mathrm{ip}}$, where $a_0^{\mathrm{ip}}$ is the in-plane pristine lattice constant.

## Supplementary References

[1] G. Kresse and J. Furthmüller, Efficient iterative schemes for *ab initio* total-energy calculations using a plane-wave basis set, Phys. Rev. B **54**, 11169 (1996).
[2] J. P. Perdew, K. Burke, and M. Ernzerhof, Generalized Gradient Approximation Made Simple, Phys. Rev. Lett. **77**, 3865 (1996).
[3] H.-R. Fuh and G.-Y. Guo, Intrinsic anomalous Hall effect in nickel: A GGA+U study, Phys. Rev. B **84**, 144427 (2011).
[4] A. A. Mostofi, J. R. Yates, Y.-S. Lee, I. Souza, D. Vanderbilt, and N. Marzari, wannier90: A tool for obtaining maximally-localised Wannier functions, Comput. Phys. Commun. **178**, 685 (2008).